# Demonstration of an air-slot mode-gap confined photonic crystal slab nanocavity with ultrasmall mode volumes

Jie Gao[*], J. F. McMillan, Ming-Chung Wu

*Optical Nanostructures Laboratory, Columbia University, New York, NY 10027 USA*

Solomon Assefa

*IBM TJ Watson Research Center, Yorktown Heights, NY 10598 USA*

Chee Wei Wong[*]

*Optical Nanostructures Laboratory, Columbia University, New York, NY 10027 USA*

**Abstract:** We demonstrate experimentally an air-slot mode-gap photonic crystal cavity with quality factor of 15,000 and modal volume of 0.02 cubic wavelengths, based on the design of an air-slot in a width-modulated line-defect in a photonic crystal slab. The origin of the high $Q$ air-slot cavity mode is the mode-gap effect from the slotted photonic crystal waveguide mode with negative dispersion. The high $Q$ cavities with ultrasmall mode volume are important for applications such as cavity quantum electrodynamics, nonlinear optics and optical sensing.

[*] Corresponding authors: jg2499@columbia.edu, cww2104@columbia.edu

Photon confinement and processes in optical cavities are critical for a vast span of fundamental studies and applications. Many optical cavities with high quality factor (*Q*) and small mode volume (*V*) have been developed and fabricated, including Fabry-Perot cavities [1], microspheres [2], silicon and silica whispering gallery type resonators [3-4] and photonic crystal nanocavities [5-6]. In particular, the two-dimensional (2D) photonic crystal cavity with a single emitter possesses remarkable possibilities towards efficient single photon sources or strong coupling regime for quantum communications and computing. Silicon photonic crystal cavities have achieved remarkable *Q* up to ~ $10^7$ experimentally but with mode volume traditionally limited to sizes that are on the order of the wavelength of light. From the definition of the normalized dimensionless effective mode volume $V_{eff} = \frac{\int \varepsilon(\vec{r}) |\vec{E}(\vec{r})|^2 d^3r}{\varepsilon(\vec{r}_{max}) \max[|\vec{E}(\vec{r})|^2]} \left(\frac{2n(\vec{r}_{max})}{\lambda}\right)^3$ (where $\vec{r}_{max}$ is the location of the maximum squared electric field), we know that mode volume can be reduced by increasing the mode maximum electric filed and localizing the mode maximum in the low index region. Reducing $V_{eff}$ in cavities enables one to control the degree of light-matter interaction for processes such as spontaneous emission, nonlinear optics and exciton-photon coupling. The Purcell factor for an emitter in a resonant cavity is inversely proportional to $V_{eff}$ [7], and the Rabi frequency is inversely proportional to the square root of $V_{eff}$ [8]. Towards this objective, Robinson and Lipson *et al.* [9] have proposed one-dimensional microcavities with ultrasmall mode volume by using dielectric discontinuities with sub-wavelength dimensions as a means of local field enhancement. Under the same idea, our previous work designed and measured an air-slot L3 cavity with ultrasmall mode volume through tapered fiber coupling [10]. However quality factors of air-slot L3 or 1D photonic crystal cavities are typically not very high because the structures suffer large vertical radiation loss at the abrupt termination of the air-slot. In this Letter, we examine mode-gap confined nanocavities in a 2D photonic crystal slab (PCS) with a non-terminated air-slot, which have both high $Q$ and ultrasmall mode volume for the study of light matter interaction, such as nonlinear phenomena [11] and cavity QED [12].

We experimentally study air-slot mode-gap photonic crystal cavities based on the original proposal in Ref. [13]. A non-terminated air-slot is added to width-modulated line-defect photonic crystal cavities (shifting air holes away from waveguide [14]) to create ultrasmall mode volume cavities. However, other work on high $Q$ heterostructure air-slot cavities [15] emphasize the positive dispersion of slotted photonic crystal waveguide mode and form the cavities by locally compressing the lattice to pull the eigenstate up in the band gap, which is opposite to conventional mode-gap cavities achieved pulling the eigenstate down in the band gap [14,16]. In order to better understand all the modes existing in the air-slot mode-gap cavities, we first investigate the modes in slotted photonic crystal waveguide and their dispersion properties. We perform three-dimensional (3D) band structure calculation for slotted W1 waveguide in Fig. 1(a). Three waveguide modes are shown in Fig. 1(b-I, II, III). Compared with the field distribution of standard W1 photonic crystal waveguide, Mode I and II can be respectively traced back to W1 waveguide fundamental even mode and high order odd mode inside the band gap [17, 18]. The electric field of mode I inside the air slot region is the Ey component and the existence of air slot enhances this component because the electric displacement in y direction needs to be continuous across the slot boundary. However the electric filed of mode II inside the air slot region is the Ex component and it is not affected since its polarization is parallel to the slot. This also explains why mode I appears close to the air band, and even has higher frequency than mode II. Note that we have mode III in the band gap, which shows a positive slope in the first Brillouin zone. While the mode is similar to the mode discussed in Ref. [19], but the origin of this mode is not well explained in the reference. By comparing the Hz field of mode III with the index-guided modes in W1 waveguide, we understand that mode III actually origins from the second index-guided mode shown in Ref. [17] below the projected bulk modes. This mode has Ey component in the slot, and so the frequency is pulled up to be within the photonic band gap.

When we generate cavities by locally shifting the air holes away from the center of waveguide, a mode-gap region is formed and cavity modes are created just below the transmission band of the slotted waveguide. There are three possible modes in the cavities, which are shown in Figure 2. Confirmed from the mode frequency and symmetry, mode in Figure 2 (a) is due to the mode gap of slotted waveguide mode I in Figure (b-I) and is expected to have both high *Q* and ultrasmall *V*. This is also the only accessible high *Q* mode in the experiment through the excitation from a single-mode strip waveguide in our design. The normal component of the electric displacement is continuous across the boundary of the air-slot walls, thus $\varepsilon_0 E_{air} = \varepsilon_{si} E_{si}$. The mode has the maximum squared electric field located in the center of the air-slot region, shown in the inset of Figure 2 (a). Compared with mode-gap confined PCS nanocavities without slot, the slot cavity $\max|\vec{E}(\vec{r})|^2$ is increased by $\varepsilon_{si}/\varepsilon_0 = n_{si}^2$. Together with $n(\vec{r}_{max})=1$, the dimensionless effective mode volume shows a total theoretical decrease of $n_{si}^5 \sim 510$ times stronger localization. Figure 2 (b) represents the mode with the same odd symmetry as mode II in slotted waveguide, and Figure 2 (c) represents the low *Q* mode which couples to photonic crystal waveguide mode III without mode-gap confinement. Note that if we form a mode gap region by moving the air holes near the defect region towards the center of waveguide, which is opposite to the conventional mode-gap confined design, we can get a localized mode just above the transmission band of slotted waveguide mode III, similar to Ref [15] with the locally compressed heterostructures.

An example scanning electron micrograph (SEM) of the electron-beam nanofabricated silicon suspended photonic crystal slot cavity is illustrated in Figure 3 (a). We fabricate the air-slot mode-gap cavities from silicon-on-insulator wafers consisting of a 220 nm thick Si layer on top of a 2μm $SiO_2$ cladding layer with electron-beam lithography and reactive-ion etch [22]. The $SiO_2$ layer is etched in buffered hydrofluoric acid to suspend the photonic crystal structures. 3D FDTD simulations were performed to numerically evaluate the properties of the fabricated cavity mode with a = 490nm, r = 0.34a, t = 0.449a, $n_{si}$=3.48, da=0.0286a, db=0.019a and dc=0.0095a. For slot width *s*=80nm (grid spacing is set to be 28nm), the air-slot mode-gap confined PCS nanocavity supports a high *Q* localized even mode (See Figure 3 (b)) with *Q* factor of $1.6 \times 10^5$ and mode volume *V* of 0.02 $(\lambda/n_{air})^3$. The *Q/V* ratio of $8 \times 10^6 (\lambda/n_{air})^3$ is much higher than previous reported ultrasmall mode volume cavity designs [9, 10, 21]. Furthermore, 2D Fourier transform of the electric field in Figure 3 (c) which shows very few leaky components inside the light cone supports the high *Q* character of this air-confined mode. To couple into the cavity from a waveguide, in-line coupling configuration was utilized [22] with cavity-waveguide distance $L_C$=4.5a. The base width of a line defect for a cavity is $\sqrt{3}$a and the width of input/output waveguides is 1.05 $\sqrt{3}$a . SEM of the designed in–line coupling sample is shown in Figure 4 (a). To probe the silicon air-slot photonic crystal cavity, we inject photons from the coupling waveguide facet and the cavity is excited via the tunneling photon. A polarization controller and a lensed fiber are used to couple transverse-electric (TE) polarized light from a tunable laser source into the coupling waveguide. The vertical radiation from the top of the nanocavities (shown as the red circle region in Figure 4 (a)) into free space is collected by a 40× objective lens (NA 0.65) and a 4× telescope, and sent to the photodetector with lock-in amplification to analyze the cavity resonances and *Q* factors obtained from the linewidth of the radiation spectrum linewidth. Figure 4 (b-d) illustrate the measured vertical radiation spectra of air-slot cavities with different slot widths (80nm and 120nm) and air hole radius (r=0.306a, 0.285a).

For the 80nm slot width devices, Figure 4 (b) shows the full width at half maximum (FWHM) from the Lorentzian fit of the radiation spectrum is of 0.13nm with a Q of ~ 12,000. The measurements are one order of magnitude smaller than the 3D FDTD simulation due to variations of the slot width and roughness of slot edge during fabrication, which degrade the cavity quality factor. For device with smaller air hole radius, the whole band gap shift towards lower frequency. Correspondingly we observe that the resonance peak shifts towards the higher wavelength in the radiation spectrum from Figure 4(b) to 4(c). When the slot width increases to 120nm, cavity mode moves up to be very close to the air band and becomes very lossy. It is expected that cavity resonance moves to shorter wavelength and *Q* factor drops significantly. However, we increase the width of the base waveguide in this case to be $1.2 \times \sqrt{3}a$. This maintains the cavity mode in the middle of the band gap, which results in *Q* of ~15,000 in our measurements shown as Figure 4(d).

This experimental demonstration of air-slot mode-gap cavity with both high *Q* and ultrasmall mode volume provides us a very good platform to advance cavity QED based on colloidal nanocrystals and silicon nanocavities in the near-infrared. Compared to InAs quantum dots, colloidal nanocrystals have the advantageous that because they can be spun on silicon nanocavities after the cavity fabrication as a post-CMOS hybrid integration process. Moreover lead salt nanocrystals can operate at the longer near- to mid-infrared wavelengths. Furthermore, in the regime where the cavity linewidth is much smaller than that of the emitter (as is the case for emitters at room temperature with large dephasing rate), in this regime the emitter *Q* ($Q_e = \omega_e / \Delta\omega_e$, where $\Delta\omega_e$ is the linewidth of the emitter) replaces the cavity *Q* in the Purcell factor $F_p = \dfrac{6Q_c}{\pi^2 V_{eff}}$ [9, 23]. Thus in this regime increasing the cavity *Q* has no effect on the radiative lifetime and the only means to increase the spontaneous emission rate is to decrease $V_{eff}$. Here we estimate the Purcell factor of air-slot mode-gap confined cavity (with $Q_c \sim 1.6\times10^5$, $V_{eff} \sim 0.02$) with PbS nanocrystals on the surface of the slot edge, and compare to that of a conventional L3 cavity (with $Q \sim 2.6\times10^5$, $V_{eff} \sim 0.7$) [24] with quantum dots positioned in the center of cavity surface plane. The single PbS nanocrystal has relatively large homogeneous linewidth due to possibly fast dephasing and spectral diffusion [25]. Even in this scenario, our air-slot mode-gap cavity still exhibits a Purcell factor of 42 while L3 cavity only has a Purcell factor of 1.2 when $Q_{emitter} = 500$. The increased Purcell factor illustrates the strong potential of the slot cavity-colloidal nanocrystal system to achieve high efficiency single photon source on-demand and even strong exciton-photon coupling in the fiber network communication frequencies.

In summary, 2D mode-gap air-slot photonic crystal cavities from the negative dispersion waveguide mode have been fabricated and experimentally characterized with $Q$ up to $\sim 15,000$ and mode volume as small as $0.02(\lambda/n_{air})^3$. Furthermore we discuss the modes existing in the slotted photonic crystal waveguides and explain in detail the origins of the air-slot mode-gap confined cavities modes. The air-slot cavity reported here can strongly enhance light-matter interactions, such as the increased Purcell factor and Rabi frequency in fundamental cavity QED studies.

The authors thank Dr. X.D. Yang and C.A Husko for helpful discussions. We acknowledge funding support from the National Science Foundation CAREER Award (NSF ECCS 0747787), DARPA MTO, and the New York State Office of Science, Technology and Academic Research.

**Figure Captions:**

Fig.1 (a) Band structure of slotted W1 photonic crystal waveguide with a = 490nm, r = 0.34a, t = 0.449a, $n_{si}$=3.48 and slot width s=80nm. (b-I,II,III) Field distributions (Left: $H_Z$; Right: $|E|^2$) of the three modes inside TE band gap.

Fig.2 2D FDTD simulation of electric intensity of cavity modes (a) (b) (c) created from photonic waveguide mode I, II, III respectively. The insets show the electric intensity distribution across the y direction (perpendicular to the slot): x=0 for mode (a, b, c) and x=a/2 for mode (c).

Fig.3 (a) SEM of the air-slot mode-gap confined cavity. (b-c) 3D FDTD simulation and Fourier transform of the electric field for cavity mode due to mode-gap effect of waveguide mode I .

Fig. 4 (a) SEM of the air-slot mode-gap confined cavity with in-line coupling waveguide. The base width of a line defect $W_0=\sqrt{3}a$ and the width of input/output waveguides is $W_c$= 1.05×$\sqrt{3}a$ . The distance between the cavity and coupling waveguide is $L_C$=4.5a. Red circle indicates the region from which the radiation signals are collected. (b-d) Radiation spectra of three devices: (b) $W_0=\sqrt{3}a$, s=80nm and r=0.306; (c) $W_0=\sqrt{3}a$, s=80nm and r=0.285; (d) $W_0$=1.2×$\sqrt{3}a$, s=120nm and r=0.285 (final fabricated radii show 2% larger than the design in this device).

**Figure 1**

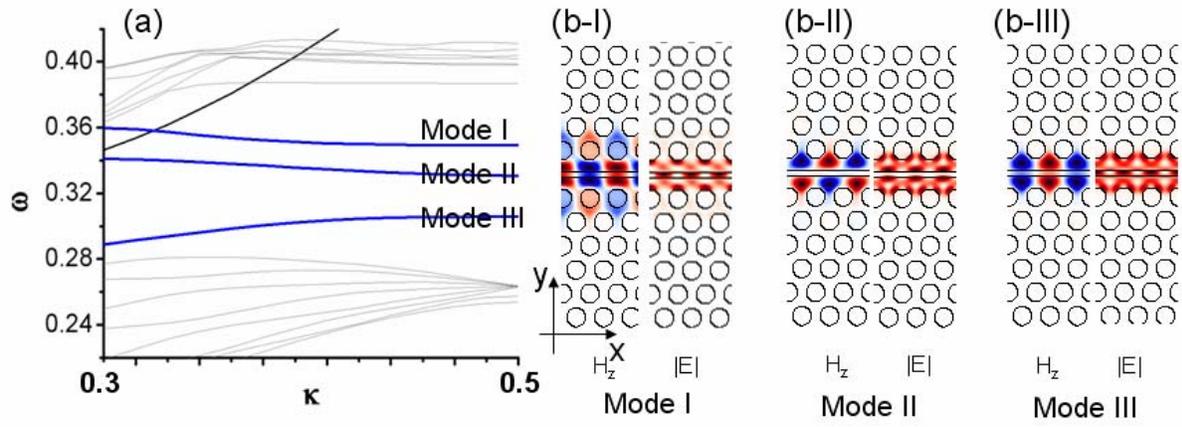

**Figure 2**

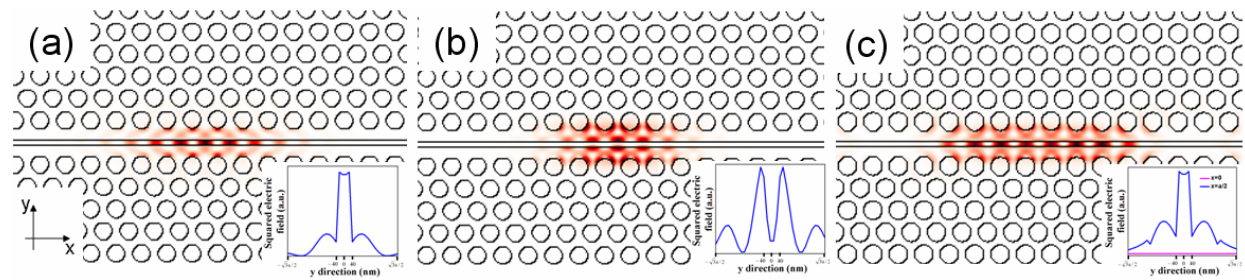

**Figure 3**

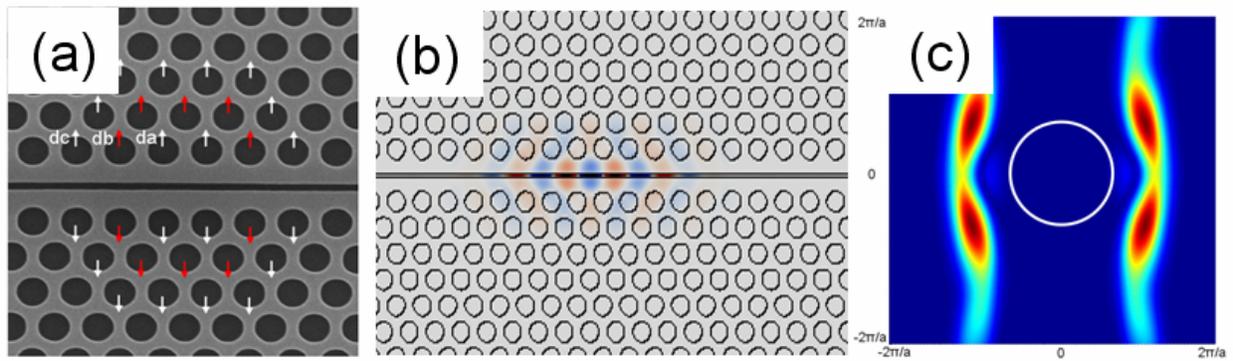

**Figure 4**

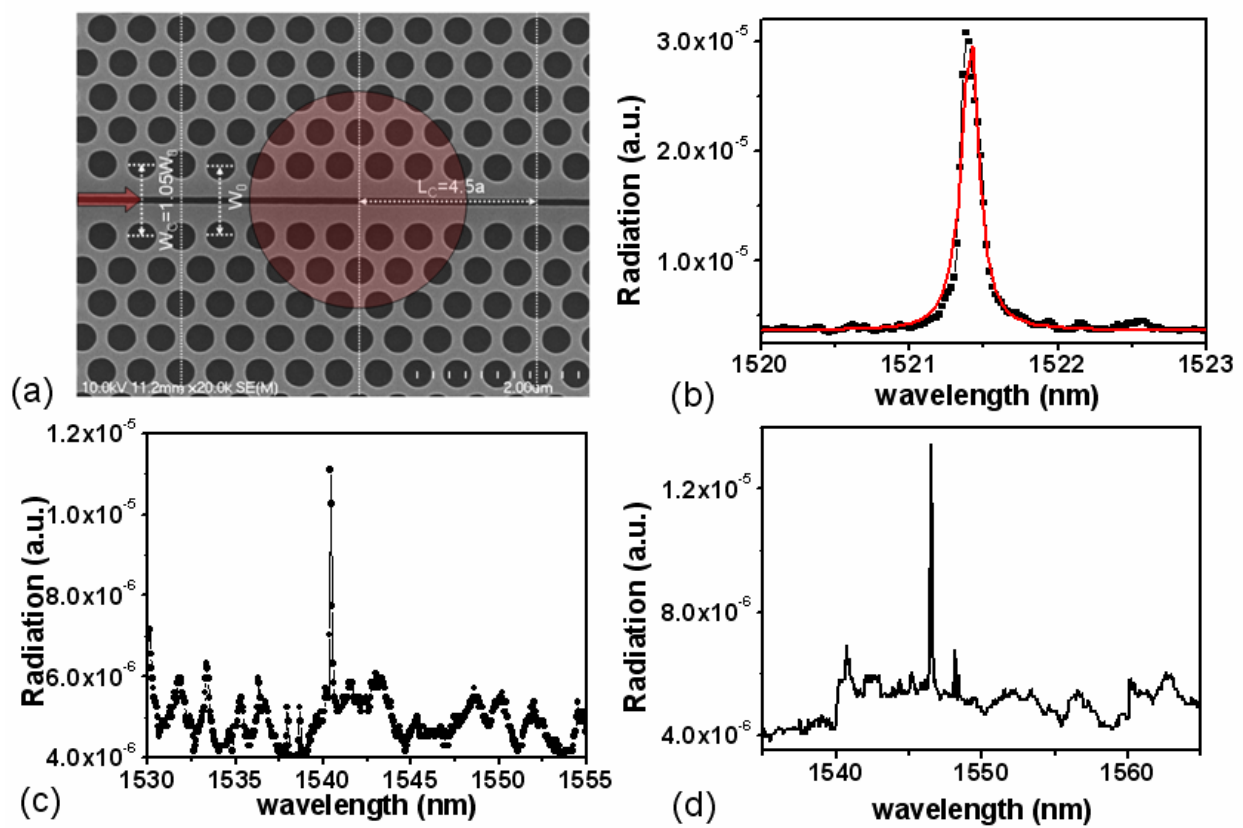

12